# Infrared Bolometers Based on 40-nm-Thick Nano-Thermoelectric Silicon Membranes

Anton Murros [a)], Kuura Sovanto, Jonna Tiira, Kirsi Tappura, Mika Prunnila, and Aapo Varpula [a)]

VTT Technical Research Centre of Finland Ltd, Tietotie 3, FI-02150 Espoo, Finland
[a)] Authors to whom correspondence should be addressed: anton.murros@vtt.fi and aapo.varpula@vtt.fi

Abstract
State-of the-art infrared photodetectors operating in the mid- and long-wavelength infrared (MWIR and LWIR) are largely dominated by cryogenically cooled quantum sensors when the target is the highest sensitivity and detection speeds. Nano-thermoelectrics provide a route towards competitive uncooled infrared bolometer technology that can obtain high speed and sensitivity, low-power operation, and cost-effectiveness. We demonstrate nano-thermoelectric LWIR bolometers with fast and high-sensitivity response to LWIR around 10 μm. These devices are based on ultra-thin silicon membranes that utilize the dimensional scaling of silicon nanomembranes in thermoelectric elements and are combined with metallic nanomembranes with subwavelength absorber structures. The fast device performance stems from a low heat capacity design where the thermoelectric beams act both as mechanical supports and transducer elements. Furthermore, by scaling down the thickness of the thermoelectric beams the thermal conductivity is reduced owing to enhanced phonon boundary scattering, resulting in increased sensitivity. The nano-thermoelectric LWIR bolometers are based on 40-nm-thick $n$- and $p$-type silicon membranes with LWIR (voltage) responsivities up to 1636 V/W and 1350 V/W and time constants in the range of 300–600 μs, resulting in specific detectivities up to $1.56 \times 10^8$ cmHz$^{1/2}$/W. We also investigate the use of a heavily doped N++ substrate to increase optical cavity back reflection, resulting in an increased Si substrate reflectance from 30% to 70%–75% for wavelengths between 8–10 μm, resulting in an increase in device responsivity by approximately 20%.

## 1  Introduction

Historically, thermal infrared (IR) detectors were restricted to low-cost and low-performance applications, due to the dominance of cryogenically cooled quantum sensors with superior sensitivities and detection speeds [1]. However, in the early 2000s it was demonstrated that uncooled thermal sensors could be utilized to achieve good thermal imaging by implementing large detector arrays [2,3]. This progress paved the way for the further study and development in the use of thermal IR sensors in imaging applications [2]. With modern advancements in uncooled thermal

IR sensing, improved device architectures, cheap silicon micromachining techniques, and improved materials, the appeal and market value of these sensors has been growing significantly, leading towards a widespread use of IR thermal sensors [4].

Thermal IR sensors find numerous applications wherein detecting or imaging the thermal emission of objects or interaction of infrared radiation with them are desired. Objects emit and absorb IR radiation characteristic of their material properties and current state, the detection of which enables applications such as infrared spectroscopic chemical analysis of chemical constituents, industrial quality control or thermal imaging for night vision and remote thermometry [5–8]. Other applications for IR imaging and spectroscopy include the detection of cancerous cells [9], thermography in medicine, biology, and sports [10,11] as well as in industrial applications such as bioprocess monitoring [12] and quality control [8].

The two most commonly available IR sensor types are quantum detectors based on electron-hole pair generation and thermal detectors (bolometers). Bolometers measure IR radiation indirectly by a two-stage signal transduction process including the conversion of incident radiation power into heat by an absorber, followed by the transduction of the thermal signal into an electrical output [5]. Quantum sensors have typically higher sensitivity and detection speed but require cryogenic cooling and exotic materials to access the long-wave IR (LWIR) range where photon energies are small. On the other hand, state-of-the-art (SoA) bolometers are less sensitive, but can still reach relatively high sensitivities without the need for cryogenic cooling. Thermoelectric bolometers [5,13–15] convert a radiation induced temperature rise into a measurable voltage via the Seebeck effect. The principle of operation of a thermoelectric bolometer is illustrated in Fig. 1(a). The bolometer comprises of a thermally isolated absorber element that is suspended by support beams. To reduce the thermal mass, optically thin absorbers coupled to an optical cavity are typically used [see Fig. 1(b)]. These kinds of structures are often called antiresonant interference structures [16], or quarter-wave resistive absorbers and Salisbury screens, when they are based on electrically conductive absorbers [17]. To maximize absorption, the effective sheet resistance of the absorber is matched to the vacuum impedance, $Z_{vac} = 377\ \Omega$ [18]. The absorber converts incident optical radiation into a temperature gradient between the absorber and the ambient (substrate), which is transduced into an electrical signal by the thermoelectric n- and p-type materials. The target wavelength of interest can be tuned by control of the depth of the optical cavity. The detector is typically sealed in a low thermal conductivity atmosphere or vacuum to increase thermal isolation.

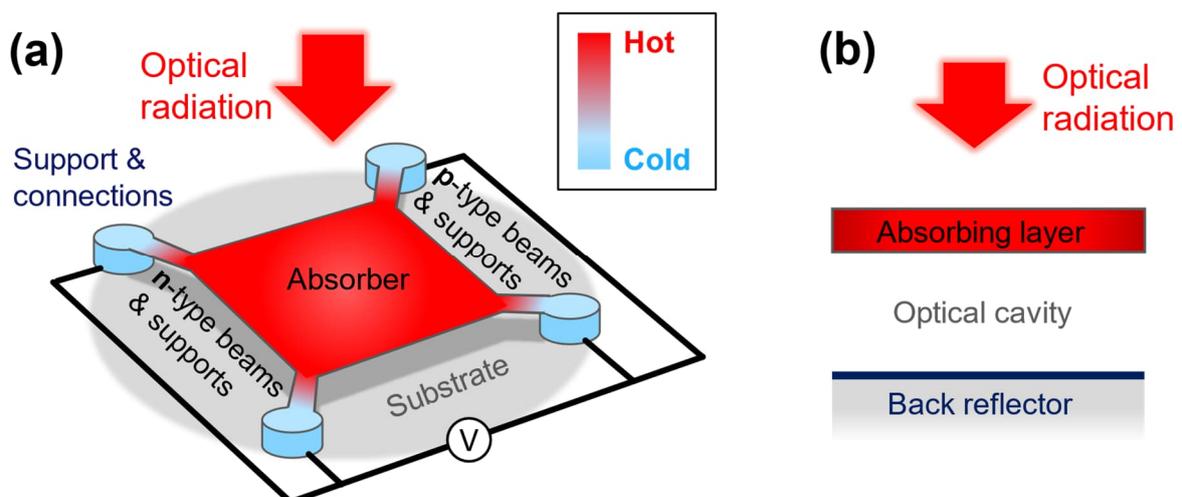

Figure 1. (a) Schematic top view of a thermoelectric bolometer illustrated by a 4-beam design utilized in this work. (b) Cross-sectional schematic of a bolometer design based on the quarter-wave absorber or an antiresonant interference structure where a thin absorber which is coupled to an optical cavity with a back reflector.

We have demonstrated nano-thermoelectric thermal devices [19] and IR bolometers [14,20–22] based on suspended ultra-thin silicon membranes. They utilize the dimensional scaling of silicon nanomembranes in thermoelectric elements in combination with metallic nanomembrane with subwavelength absorber structures, and integrated device design to produce a low thermal mass absorber. The thermoelectric beams, which act as both mechanical supports and transducer elements, preclude the need for any additional micromechanical supports, reducing the thermal mass and additional thermal link of the detector to the environment. This enables high speed and high sensitivity operation. Silicon is an attractive material for nano-thermoelectrics, since it has been demonstrated that in silicon membranes reducing the thickness of the membrane to the order of 100 nm and below results in a proportional and significant reduction in thermal conductivity [23–27]. Experiments have shown that by reducing the thickness of silicon thin films the thermal conductivity can be reduced by almost two orders of magnitude, with an experimental minimum thermal conductivity of 9 W/mK measured in a 9 nm free-standing high crystalline quantity silicon membrane [23]. The reduction in thermal conductivity can be attributed to reduced phonon transport properties owing to spatial confinement of phonon modes in the thin film [28–30] and increased boundary scattering, resulting in a shortened phonon mean free path [31–33]. This decrease in thermal conductivity results in improved thermoelectric detector performance as it increases the thermal isolation and the thermoelectric figure of merit $ZT$ of the material, leading to increased sensitivity and signal-to-noise ratio [14,19].

By applying advanced micromachining techniques, absorber and device designs with low-thermal mass [14,20,34], nanoantenna absorbers [35], conventional high-performance thermoelectric materials [36,37] and nano-thermoelectrics [14,20,21,38], thermal IR sensors have been shown to be relatively competitive to SoA quantum detectors. In fact, the fundamental limits for sensitivity of uncooled thermal IR sensors operated at ambient temperatures in the LWIR region of 8–15 μm are quite close to quantum sensors, with thermal sensors being favoured at longer wavelengths and higher operating temperatures [2]. Interestingly, antenna-coupled thermoelectric bolometers have also been used in THz detection [39]. To improve the performance further, control of the thermal properties of the detector, for example in terms of thermal mass and thermal isolation are essential.

In this article, we report for the first time nano-thermoelectric bolometers based on 40 nm poly-silicon (poly-Si) nanomembranes. Their specific detectivities $D^*$ and time constants τ are competitive with commercial and non-commercial LWIR SoA bolometers. We demonstrate thermoelectric IR bolometers with poly-Si thermoelectric membranes with a reduced thickness of 40 nm from 70–80 nm of the previous works [14,21], and investigate the use of heavily doped N++ substrate to increase optical cavity back reflection, showing an improvement in device IR responsivity. We obtain IR (voltage) responsivities up to 1636 V/W and 1350 V/W with 200°C and 800°C blackbody radiation, respectively, and time constants in the range of 300–600 μs, resulting in specific detectivities up to $1.56 \cdot 10^8$ cmHz$^{1/2}$/W. We show that the N++ substrate increases the device responsivity by ~20%. We discuss the effect of tuning the grid geometry of the membrane on the speed and responsivity of the devices. The bolometers we demonstrate are scalable to various matrix and pixel sizes. These results demonstrate the scalable nature of nano-thermoelectrics based IR detection, enabling uncooled IR detection technologies that approach SoA quantum detectors.

## 2 Model of thermoelectric bolometers

### 2.1 Device model

Thermoelectric bolometers can be described using a model based on a lumped thermal RC circuit [14,19,21,40]. The thermoelectric beams of the present bolometers act as both supporting structures and transducer elements to transform the temperature difference between the surrounding bath and the absorber into a measurable voltage. This voltage, $V$, is characterized by the total Seebeck coefficient, and is given by $S = dV/dT = S_p - S_n$, where $T$ is the absolute temperature, and $S_p$ and $S_n$ are the Seebeck coefficients of the p- and n- type thermoelectric elements, respectively. The speed of the bolometer is described by the time constant, given by $\tau = C_{th}/G_{th}$, where $C_{th}$ is the heat capacity of the absorber and $G_{th}$ is the thermal conductance of the bolometer to the substrate and surroundings. $G_{th}$ is comprised of two heat transfer mechanisms – conduction and radiation – and is given by $G_{th} = G_\kappa + G_R$. Conductive heat transfer, $G_\kappa = (\kappa_p + \kappa_n)t_{TE}/N$, can be described using the sum of the thermal conductivities of the n- ($\kappa_n$) and p- type ($\kappa_p$) thermoelectric elements, respectively, assuming that the thickness, $t_{TE}$, and number of squares in series (i.e. length-to-width ratios), $N$, of the n- and p-type thermoelectric elements are equal. The devices of this work consist of 2 n-type thermoelectric beams and 2 p-type thermoelectric beams in parallel. The number of squares in series in a single beam is $N_{beam} = L/W$, where $L$ and $W$ are the length and width of the beam. This leads to $N = N_{beam}/2$. Radiative heat transfer is given by $G_R = 4A_{abs}\epsilon\sigma_{SB}T^3$, where $A_{abs}$ is the area of the absorber, $\epsilon$ is the emissivity, and $\sigma_{SB}$ is the Stefan-Boltzmann constant.

The total optical efficiency of a bolometer depends on the spectrum of the incident power, as the wavelength of incident radiation has an effect on the optical efficiency of absorbers. This can be described by the wavelength-dependent optical (spectral) efficiency $\eta(\lambda)$ of the absorber, which determines the absorbed power at the wavelength $\lambda$. The total optical efficiency can then be written as the ratio between the absorbed optical power $P_{abs}$ and the total optical power $P$ incident on the absorber as $\eta_{tot} = \frac{P_{abs}}{P} = \frac{\int \eta(\lambda)P_\lambda(\lambda)d\lambda}{\int P_\lambda(\lambda)d\lambda}$, where $P_\lambda(\lambda)$ is the spectral incident optical power at wavelength $\lambda$. In general, the optical characteristics of the absorber can be modelled with finite element methods (see e.g. Ref. [14]) or in some cases with the transfer matrix method or even an analytical model [22].

The frequency dependence of the output voltage amplitude of a thermoelectric bolometer is given by [14]

$$V = \frac{V_{ampl}}{\sqrt{1 + \tau^2\omega^2}},\tag{1}$$

where $V_{ampl} = S\eta_{tot}P/G_{th}$, and $\omega$ is the angular frequency of the optical power. The phase of the output voltage is

$$\theta = \arctan(-\tau\omega).\tag{2}$$

Below the cut-off angular frequency of $\omega_c = 1/\tau$, Eq. (1) can be simplified to give $dV/dP = S\eta_{tot}/(G_{th}\sqrt{1 + \tau^2\omega^2})$ and the voltage responsivity becomes

$$R_V = \frac{dV}{dP}\bigg|_{\omega \ll \omega_c} = \frac{S\eta_{tot}}{G_{th}}.\tag{3}$$

With $\tau = C_{\text{th}}/G_{\text{th}}$ Eq. (3) can be written in a form which displays the trade-off between the sensor speed ($\tau$) and responsivity ($R_V$):

$$R_{\text{V}} = S\eta_{\text{tot}} \times \frac{\tau}{C_{\text{th}}}. \tag{4}$$

To show the dependencies of $R_V$ and $\tau$ on all material and geometrical parameters, we utilize a model describing the devices of this work [22,41]. With $G_{\text{th}} \approx G_\kappa = (\kappa_{\text{p}} + \kappa_{\text{n}})t_{\text{TE}}/N$, Eq. (3) can be written as

$$R_{\text{V}} = \frac{S\eta_{\text{tot}}}{\kappa_{\text{p}} + \kappa_{\text{n}}} \times \frac{N}{t_{\text{TE}}}. \tag{5}$$

The absorber heat capacity is a sum of the heat capacities of the thermoelectric and absorber materials, $C_{\text{th}} = A_{\text{abs}}^{\text{mat}}( c_{\text{V,abs}}t_{\text{abs}} + c_{\text{V,TE}}t_{\text{TE}})$, where $A_{\text{abs}}^{\text{mat}}$ is the total area occupied by absorber material, $c_V$ is the volumetric heat capacity of the thermoelectric element, $c_{\text{V,abs}}$ is the volumetric heat capacity of the absorber material, and $t_{\text{abs}}$ is the thickness of the absorber material. With the above $C_{\text{th}}$, the formula for the thermal time constant becomes

$$\tau = \frac{NA_{\text{abs}}^{\text{mat}}}{(\kappa_{\text{p}} + \kappa_{\text{n}})t_{\text{TE}}}( c_{\text{V,abs}}t_{\text{abs}} + c_{\text{V,TE}}t_{\text{TE}}), \tag{6}$$

which shows that the time constant is directly proportional to the geometric design parameters $N$ and $A_{\text{abs}}^{\text{mat}}$, proportional to the specific heat capacities $c_{\text{V,abs}}$ and $c_{\text{V,TE}}$, and the thickness of the absorber material $t_{\text{abs}}$, and inversely proportional to $t_{\text{TE}}$. By scaling down the thickness, $t_{\text{TE}}$, of the thermoelectric elements, the thermal mass and $G_{\text{th}}$ of the bolometer are reduced, resulting in greater temperature change per unit of incident radiative power and therefore larger responsivity [see Eq. (5)]. However, this will also reduce the speed of the bolometer [see Eq. (6)], but this can be compensated by altering the geometry of the supports, for example in terms of $N$.

The noise equivalent power (NEP) of a radiation sensor is the equivalent incident radiation power that would yield a signal-to-noise ratio of unity at the sensor's output defined here for a unit bandwidth of 1 Hz. The two principal sources of noise contributing to the total NEP of a thermal sensor include thermal fluctuation noise arising from random energy exchange between the absorber and surrounds by different heat carriers including electrons, phonons and photons, and Johnson-Nyquist noise originating from the total electrical resistance $R$ of the transducers. The optical NEP of the thermal fluctuation noise is given by

$$\text{NEP}_{\text{th}} = \frac{\sqrt{4k_B T^2 G_{\text{th}}}}{\eta_{\text{tot}}}, \tag{7}$$

where $k_B$ is Boltzmann's constant. Below the thermal cut-off frequency ($\omega \ll \omega_{\text{c}}$) the optical NEP of Johnson-Nyquist noise can be simplified to

$$\text{NEP}_{\text{JN}} = \frac{\sqrt{4k_B TR}}{R_{\text{V}}}. \tag{9}$$

These two independent noise sources can be combined into total optical NEP given by [14,19,41]

$$\mathrm{NEP} = \sqrt{\mathrm{NEP_{th}^2 + NEP_{JN}^2}} = \mathrm{NEP_{th}}\sqrt{1 + \frac{1}{\widetilde{ZT}}}, \tag{10}$$

where $\widetilde{ZT} = S^2 T/(G_{\mathrm{th}} R)$ is the effective thermoelectric figure of merit. When the geometries and the absolute values of the material parameters of the n- and p-type thermoelectric elements are equal $\widetilde{ZT}$ coincides with the material thermoelectric figure of merit $ZT = \sigma S'^2 T/\kappa$, where $\sigma$ is the electric conductivity, $S'$ the Seebeck coefficient, and $\kappa$ the thermal conductivity of the material [19]. Eq. (10) shows that above $\widetilde{ZT} > 1$ the thermal fluctuation noise dominates, and below $\widetilde{ZT} < 1$ the Johnson-Nyquist noise dominates (note that both contributions are equal when $\mathrm{NEP} = \sqrt{2}\mathrm{NEP_{th}}$). NEP can be minimized by maximizing $\widetilde{ZT}$ and $\eta_{\mathrm{tot}}$, and by minimizing $G_{\mathrm{th}}$ and $R$. These parameters can be optimized by affecting the geometries and improving the material properties – in terms of electrical and thermal conductivity, and Seebeck coefficients – of the thermoelectric elements [19]. Finally, the sensitivity of the detector can be described by the specific detectivity, defined as

$$D^* = \sqrt{A_{\mathrm{abs}}}/\mathrm{NEP}. \tag{11}$$

## 2.2   Optical absorber model

In this work we use an analytical and a semi-analytical models of an ideal quarter-wave resistive absorber to estimate the absorber [21,22] which consists of an optically thin resistive absorber placed above a back reflector. The semi-analytical model is based on the transfer-matrix method [22]. The wavelength dependence of the absorptance of the quarter-wave resistive absorber is given by [22]

$$A = \frac{4\dfrac{Z_{\mathrm{vac}}}{R_{\mathrm{sh}}}\left[\sin\left(\dfrac{2\pi d_{\mathrm{cav}}}{\lambda}\right)\right]^2}{\dfrac{Z_{\mathrm{vac}}}{R_{\mathrm{sh}}}\left(\dfrac{Z_{\mathrm{vac}}}{R_{\mathrm{sh}}} + 2\right)\left[\sin\left(\dfrac{2\pi d_{\mathrm{cav}}}{\lambda}\right)\right]^2 + 1}, \tag{12}$$

where the vacuum impedance or the impedance of free space is $Z_{\mathrm{vac}} \approx 376.73\ \Omega$, $R_{\mathrm{sh}}$ is the (effective) sheet resistance of the absorber, and $d_{\mathrm{cav}}$ is the depth of the cavity. The absorptance spectra of exemplary quarter-wave resistive absorbers are plotted in Fig. 2.

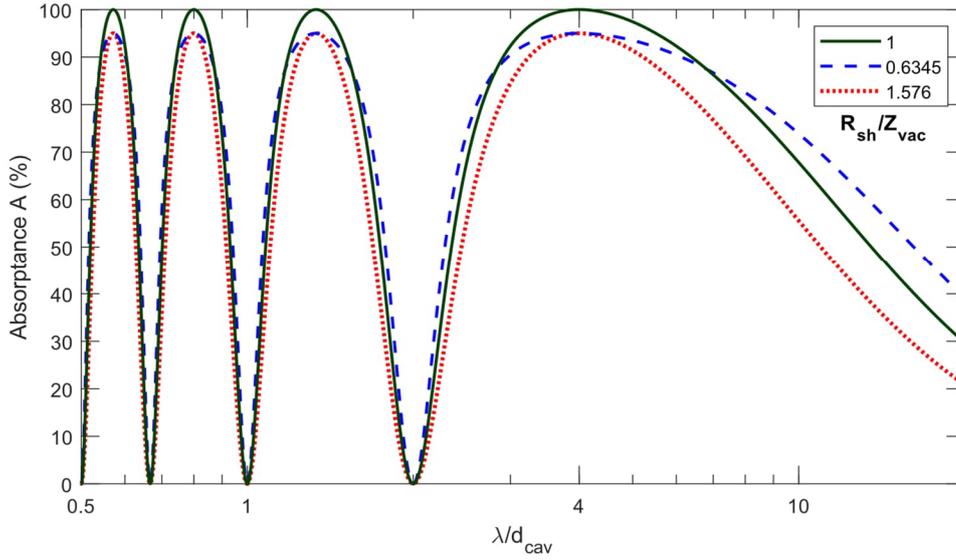

Figure 2. Absorptance spectra of ideal quarter-wave resistive absorbers with various levels of the matching of absorber optical impedance ($R_{sh}/Z_{vac}$). They were calculated using the analytical model [Eq. (12)]. The wavelength $\lambda$ is normalized with the depth of the optical cavity $d_{cav}$.

Fig. 2 and Eq. (12) shows that the absorption spectrum exhibits periodically repeating peaks where the absorption reaches maximum. The wavelengths of these peaks are given by

$$\lambda_{\text{peak}} = \frac{4 d_{\text{cav}}}{1 + 2n},$$

(13)

where the order $n$ = 0, 1, 2, 3, .... As the graph shows, the relative full-width-at-half-maximum (FWHM) bandwidth of the peak with the highest wavelength ($n$ = 0) is rather broad: 280 % [22].

If the optical impedance of the absorber is not matched to the vacuum impedance, i.e. $R_{sh} = Z_{vac}$, the shape of the absorption spectrum remains relatively unchanged (see Fig. 2), but the peak maxima are reduced. The value of the peak maxima can be calculated using [21]

$$A_{\text{peak}} = \frac{4 \dfrac{Z_{\text{vac}}}{R_{\text{sh}}}}{\dfrac{Z_{\text{vac}}}{R_{\text{sh}}} \left( \dfrac{Z_{\text{vac}}}{R_{\text{sh}}} + 2 \right) + 1},$$

(14)

which has been obtained from eq. (12). Eq. (14) is plotted in Fig. 3, showing that a rather high absorption is obtained with a wide range of sheet resistance $R_{sh}$ (for example for $A$ > 50%, $R_{sh}$ can range between 65 and 2196 Ω [22]).

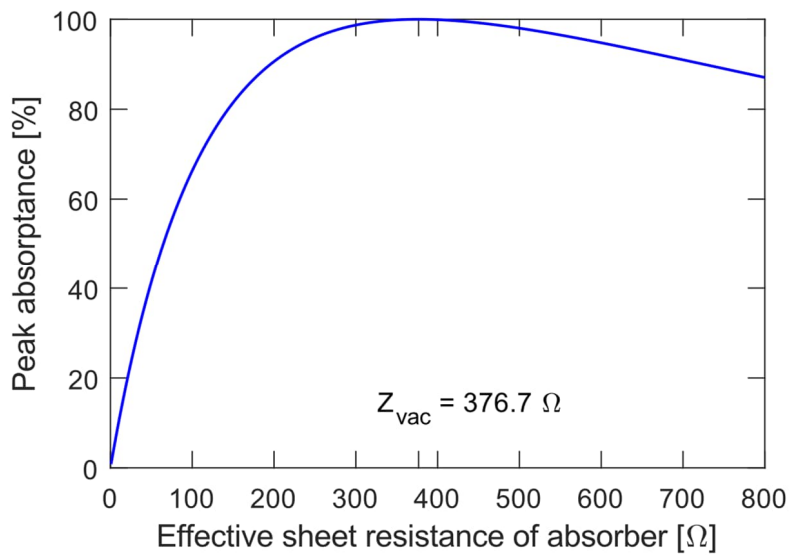

Figure 3. Dependence of peak absorptance of a quarter-wave resistive absorber $A_{peak}$ as a function of effective sheet resistance of the absorber $R_{sh}$. Calculated using eq. (14). Adapted from Ref. [21].

In the above analysis, the reflectance of the back reflector of the quarter-wave resistive absorber is assumed to be 100%. The cases with lesser reflectance are plotted in Fig. 4. They were calculated semi-numerically using the transfer-matrix method [22].

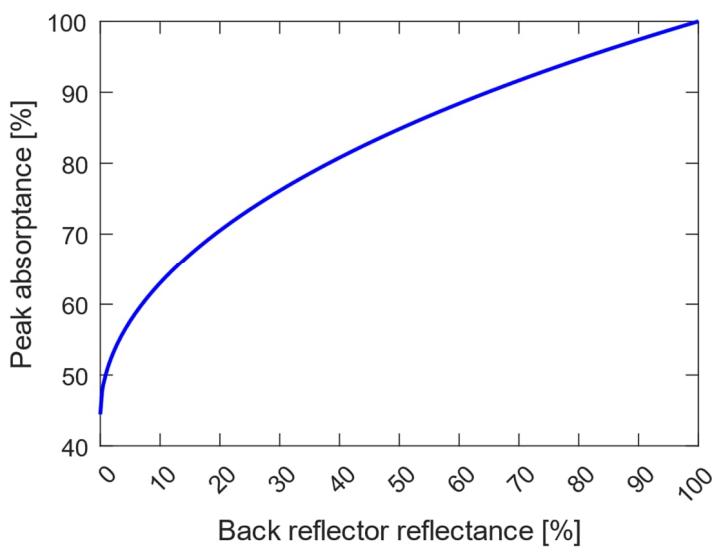

Figure 4. Dependence of the peak absorptance of a quarter-wave resistive absorber $A_{peak}$ on the reflectivity of the back reflector. Although this was calculated using the transfer-matrix method with 10 nm thin absorber with $R_{sh} = Z_{vac}$ and cavity depth $d_{cav} = 2500$ nm at wavelength $\lambda = 10$ µm, the result can be generalized to other cases with optically thin absorber which are dominantly resistive. Adapted from Ref. [22].

In addition to the analytical and semi-analytical models presented above, a full-wave electromagnetic model based on the finite element method (FEM) was developed for more detailed simulations. In these calculations, the three-dimensional Maxwell equations are solved using Comsol Multiphysics® [42]. The FEM simulations include the realistic layer stack consisting of the absorber, membrane, cavity, and substrate with perfectly matched layer (PML) boundary conditions. Periodic boundary conditions are applied in the lateral directions around the unit cell that defines the simulation volume. The incident radiation is modelled as a plane wave arriving perpendicular to the absorber. The power absorbed in the absorber is calculated based on the time-average spectral power density that is proportional to the electric field intensity and the imaginary part of the permittivity of the absorbing material as described in [43]. The permittivity of TiN and Si was determined using the Drude model with parameters similar to those in reference [14].

# 3 Experimental

## 3.1 Device fabrication

Scanning electron micrographs and cross-sectional schematics of the thermoelectric bolometers are shown in Fig. 5. For a full description of the layer structure and fabrication of nano-thermoelectric bolometers, see Ref. [14]. The bolometer consists of an optical absorber supported by 40 nm thick n- and p-type doped poly-Si thermoelectric legs, which transduce radiation induced heating in the absorber into a measurable voltage. The speed and sensitivity of the bolometer can be tuned by controlling the width of the thermoelectric legs, which defines the degree of thermal isolation. The stress of the membrane is controlled by a stress compensating frame of 50 nm $Al_2O_3$ described in a previous work [44]. The absorbing element is 24 nm reactively sputtered TiN, which also acts to electrically connect the n- and p-type poly-Si. The bolometer is patterned into a grid in order to allow for control of the impedance of the absorber, and to facilitate etching of the underlying sacrificial oxide to release the membrane. The absorber is suspended over a 2.5 μm deep optical cavity, which allows for tuning the target wavelength of interest. In this case the cavity depth was selected to maximize detector output for room-temperature thermal radiation with a target wavelength of 10 μm. The substrate acts as a back reflector and with the absorber these form a quarter-wave resistive absorber. For some devices the silicon substrate back reflection was enhanced by heavily doping the substrate surface.

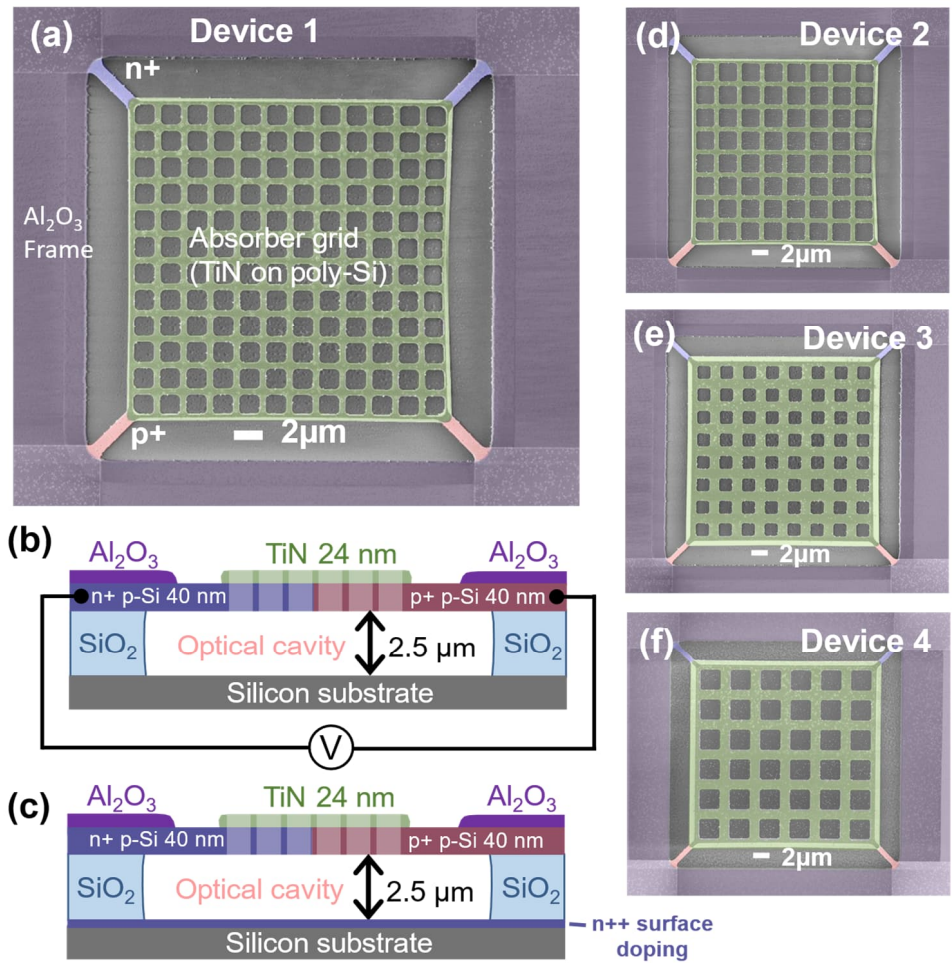

Figure 5. (a) A coloured scanning electron micrograph (top view) of a suspended bolometer with thermoelectric beams supporting a metal grid absorber. Diagonal cross-sectional schematics of the layer structure of the bolometer suspended over an optical cavity with (b) standard Si substrate and (c) N++ surface doping to form a back-reflector. (d)-(f) Coloured SEMs of Devs. 2, 3 & 4.

The devices were fabricated in the VTT Micronova cleanroom facilities on standard 150 mm single-side-polished *p*-type (1–50 Ωcm) silicon wafers following mainly the fabrication process described in Ref. [14]. A set of wafers was implanted with phosphorus and subsequently annealed at 1050°C in 2% $O_2$ for 1 hour to form an N++ surface doped back reflector on the silicon substrates. Next, a 2.5μm thick layer of sacrificial $SiO_2$ was deposited on the wafers by low pressure chemical vapor deposition (LPCVD) using tetraethyl orthosilicate (TEOS). After the TEOS deposition, the N++ surface doped wafers had sheet resistance of 4.5 Ω/sq. 40 nm of poly-Si was deposited using LPCVD and then doped selectively with boron and phosphorus by ion implantation and annealed at 700°C for 2 h. Resistivities of 8.0 mΩcm and 4.9 mΩcm, charge-carrier mobilities of 8.9 $cm^2$/Vs and 7.6 $cm^2$/Vs, and Hall carrier concentrations of 8.8 x $10^{19}$ $cm^{-3}$ and 1.7 x $10^{20}$ $cm^{-3}$ were measured for the n- and p-type 40 nm thick poly-Si, respectively. In reducing the deposited poly-Si thickness to 40 nm, the deposited film exhibits a high number of island-type growth bumps, and as a result high topography (which can be seen in Fig. S1 in the Supplementary Information). The defect density of ~40 nm growth-bumps in the 40 nm poly-Si film is 8.6 defects/$μm^2$. The poly-Si was patterned to form the thermoelectric elements, after which the stress compensation frame of 50 nm $Al_2O_3$ was deposited by atomic layer deposition (ALD) and patterned. The absorber and contact metal of 24nm TiN with sheet resistance 70 Ω/sq (electric conductivity 6.0·$10^5$ S/m) was reactively sputtered in nitrogen atmosphere, followed 500 nm of aluminium to form the device pads. n- and p-type poly-Si-TiN contact resistivities of 4350 Ω$μm^2$ and 5860 Ω$μm^2$, respectively, were measured using Cross-Bridge

Kelvin (CBK) resistors with a 4×4 µm² contact area. Finally, the devices were suspended by selective etching of the sacrificial oxide by HF vapor etching, simultaneously forming the optical cavity.

The device parameters of the bolometer devices depicted in Fig. 5 are listed in Table 1. Reflectances of 70% at 8 µm and 75% at 10 µm of the N++ surface doped substrates (with the sacrificial SiO₂ layer removed) were measured with a Semilab spectroscopic ellipsometer SE-2000 at the angle of incidence of 72 degrees. To investigate the effect of device design parameters on the device performance, each bolometer has differences in the supporting beam dimensions and absorber grid dimensions. Included in Table 1 are the resulting effective absorber impedance of each grid geometry, which should be matched to the vacuum impedance of $Z_{vac}$ = 377 Ω for maximum absorptance. Optical impedance matching is obtained by tuning the geometry of the absorber grid or thickness of the absorbing layer. The effective absorber sheet resistances of Devs. 1, 2, and 4 are 320 Ω, 396 Ω and 237 Ω, respectively, resulting in estimated absorptances of 72%–93% with the effect of the back reflection taken into account (assuming that the effective absorber optical impedance would be similar to the effective sheet resistance).

Table 1. Characteristics of fabricated nano-thermoelectric infrared bolometers.

| Device | 1 | 2 | 3 | 4 |
|---|---|---|---|---|
| Stress compensation frame | 50nm Al₂O₃ | | | |
| Absorbing material | 24nm TiN | | | |
| Si substrate surface (back reflector) | Boron-doped 1–50 Ohmcm | N++ doped | | Boron-doped 1–50 Ohmcm |
| Thermoelectric poly-Si | 40nm Poly-Si | | | |
| Thermoelectric beam linewidth [µm] | 0.9 | 0.9 | 0.6 | 0.6 |
| Grid pitch $p$ [µm] | 2 | 3 | 3 | 4 |
| Grid linewidth $w$ [µm] | 0.4 | 0.5 | 1.2 | 1.2 |
| Device area [µm²] | 30x30 | | | |
| Absorber area [µm²] | 23.8 x 24.1 | 24.0 x 24.3 | 25.1 x 24.9 | 25.0 x 24.7 |
| Total area occupied by absorber material $A_{abs}^{mat}$ [µm²] | 247 | 211 | 428 | 348 |
| Effective absorber sheet resistance[a] [Ω] | 320 | 396 | 175 | 237 |
| Back reflector reflectance | 30% | 75% | 75% | 30% |
| Peak absorptance estimated by semi-analytical model[b] | | | | |
| Experimental effective impedance and 100% back reflection[c] | 99.3% | 99.9% | 86.6% | 94.8% |
| Perfectly matched absorber with | 76% | 93% | 93% | 76% |

| experimental back reflectance[d] | | | | |
|---|---|---|---|---|
| Experimental absorber and reflector[e] | 75% | 93% | 81% | 72% |

[a]Calculated with the formula: $p/w \times$ absorber metal sheet resistance (70 Ω/sq)
[b]See Section 2.2
[c]Estimated using Eq. (14) given in Ref. [21] and Section 2.2. Assumes optimal wavelength and use of 100% back reflector on the bottom of the cavity.
[d]See Ref. [22], Section 2.2, and Fig. 3.
[e]Calculated using: Estimated absorptance with 100% back reflection × Ideal absorptance with given back reflectance.

## 3.2   Infrared characterization

The fabricated detectors were characterized optically within a vacuum chamber at below 1 Pascal pressure, and the detector temperature was kept constant with a temperature-controlled sample holder during the characterization measurement. The measurement input radiation was fed into the chamber through a ZnSe window on the vacuum chamber. In addition, the signal was optically filtered through a high-resistivity (>5 kΩcm) double-side-polished silicon wafer to ensure that none of the measured signal would be caused by the silicon photovoltaic effect. Silicon is transparent at wavelengths above ~1.1 μm, and the combined transmission by the Silicon and ZnSe filters is relatively flat in the wavelength range of 1-20 μm that was used in the characterization of the devices.

A blackbody infrared source (CI-Systems, model SR-200 33, emissivity 0.99 ± 0.01, temperature accuracy ±2 K) was used to illuminate the detectors. The optical power was controlled through the blackbody IR source temperature and by the diameter of its output aperture. The blackbody IR source temperature determines the spectrum of the radiation that reaches the detector surface. Higher initial blackbody radiation temperatures naturally result in higher signal responses. Furthermore, the created IR spectra peak at lower wavelengths for higher initial temperatures. The detectors presented in this paper have their optimal absorption range at the higher wavelength range, and therefore a low initial blackbody temperature of 200 °C was used for their power response characterization.

The detector response was measured by an SR865A lock-in amplifier and a SR560 preamplifier, to which the detector output was differentially connected. The SR865A was locked into the frequency of an optical chopper, which was used to modulate the input signal from the blackbody IR source. When characterizing the frequency response of the bolometers, the blackbody IR source temperature was set to its maximum of 1200 °C to ensure a sufficient signal level, as the input signal aperture diameter was restricted by the width of the optical chopper wheel slits. The response level was measured between the n+ and +p sides of the bolometer, where the two thermocouples were connected in parallel.

# 4   Results and discussion

The measured frequency and power responses of the fabricated bolometers are shown in Fig. 6. The thermoelectric bolometer model discussed in section 2 was fitted against the experimental data and the fitted parameter values and other device characteristics are tabulated in table 2. There is good agreement between the experimental data and fitted model, verifying the functionality of the devices.

As demonstrated in Fig. 7, the devices have a linear response across the whole scale of the incident input signal powers, with a higher response level at the lower blackbody temperature (200 °C), as was expected by their optimization to the peak of the blackbody radiation. In both the frequency and power response graphs, we can observe that Dev. 4 provides an overall higher signal level, followed by Devs. 1 and 2, while Dev. 3 shows the lowest overall response. Regardless, all the devices demonstrate expected behaviour, and the fitted curves settle easily within the margins of error.

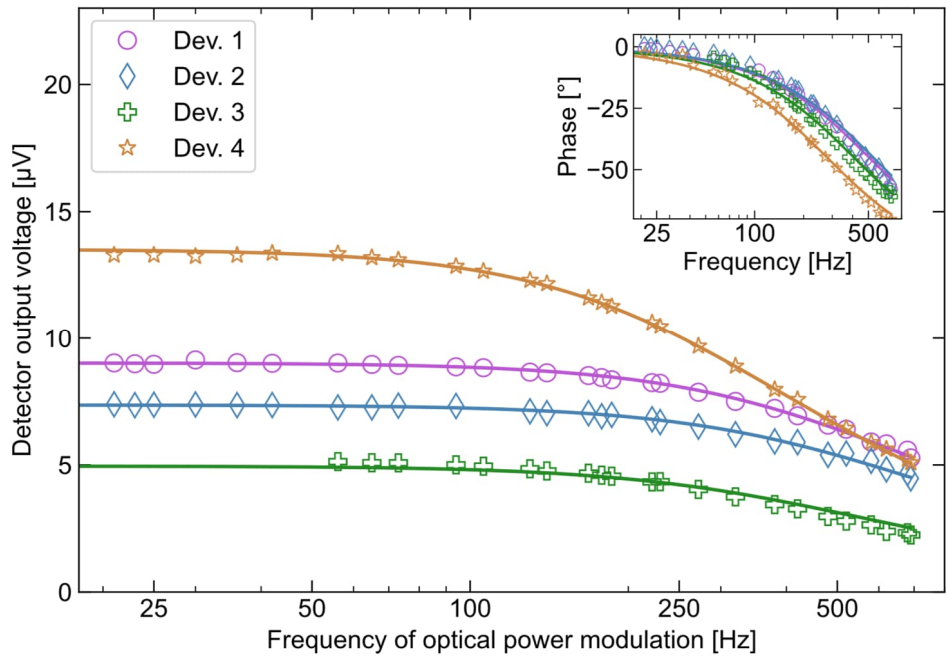

Figure 6. The measured magnitude and phase (inset) of the frequency responses of bolometers. The lines show the calculated model data using Eqs. (1) and (2) against the detector output voltage amplitudes as a function of the frequency of optical modulation. The resulting time constants of each device are tabulated in Table 2. The other fitting parameters are $V_{ampl}$ = 29.932 ± 0.006 μV, 40.440 ± 0.016 μV, 22.693 ± 0.005 μV, and 51.378 ± 0.011 μV. The devices were measured using blackbody infrared source at 800°C, due to the higher optical power required during frequency characterization.

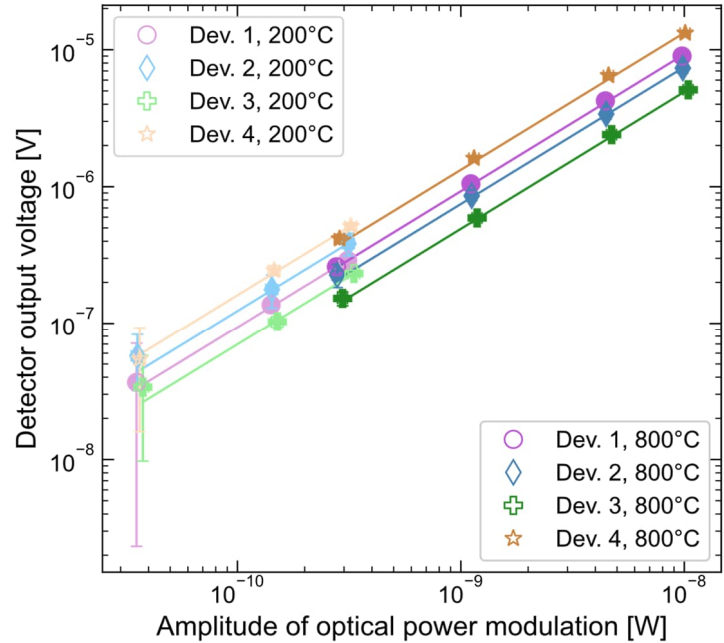

Figure 7. Power responses of bolometers measured at 200°C and 800°C with various output apertures of the blackbody infrared source. The lines show linear fits to the data that correspond with the responsivities listed in Table 2.

Table 2 lists the fitted parameter values for each of the devices. We observe responsivities up to 1636 V/W and 1252 V/W in Devs. 4 and 2, respectively, and time constants from 572 μs down to 297 μs. Owing to the increase in the thermal isolation of the devices caused by the dimensional scaling of the thermoelectric legs, these devices are not as fast as those we reported previously (66-208 μs) [14,21], however the responsivities of the devices are significantly improved. In addition, the speed of the current detectors is comparable or better than previously reported uncooled devices with $D^*$ in the same range, such as a thermoelectric detector with time constant of 560 μs [45] and resistive bolometers with time constants 170–220 μs [46], 530 μs [47], and 1500 μs [48] (see also a comparison graph in Ref. [14]).

The trade-off between device speed and responsivity can be visualized by plotting these parameters against each other, shown in Fig. 8. Fig. 8 includes data points from previous works [14,21] for comparison. Dev. 4 has very high responsivity owing to the narrower thermoelectric beams and adequate absorber effective sheet impedance matching. Dev. 3 illustrates the effect of poor impedance matching, which acts to reduce the absorptance of the quarter-wave resistive absorber. Increased thermoelectric beam linewidth acts to decrease the thermal time constant (Devs. 1 & 2), however the thermal mass of the absorber also plays a role in how quickly the device thermalizes.

Table 2. Summary of the characteristics of the nano-thermoelectric IR bolometers of Fig. 5.

| Device | 1 | 2 | 3 | 4 |
|---|---|---|---|---|
| Measured thermal time constant $\tau$ [μs] | 315.5 ± 2.2 | 297.4 ± 1.0 | 391.1 ± 5.9 | 571.5 ± 66.4 |
| Total bolometer resistance $R_{tot}$ [kΩ] | 56.3 | 121.8 | 13.6 | 40.9 |
| Resistance of thermoelectric beams $R_{beams}$ [kΩ] | 7.6 | 7.6 | 11.5 | 11.5 |
| Measured responsivity $R_V$ (200 °C blackbody radiation) [V/W] | 935.8 ± 13.1 | 1232.0 ± 28.1 | 698.3 ± 15.2 | 1609.8 ± 22.2 |
| Measured responsivity $R_V$ (800 °C blackbody radiation) [V/W] | 926.6 ± 7.5 | 747.6 ± 3.5 | 492.4 ± 4.0 | 1328.7 ± 23.4 |
| Optical noise equivalent power[a,b] (NEP) [pW/Hz$^{1/2}$] | 32.4 ± 0.5 | 36.2 ± 0.9 | 21.4 ± 0.5 | 16.1 ± 0.3 |
| Specific detectivity[a] $D^*$ [$10^7$ cmHz$^{1/2}$/W] | 7.5 ± 0.2 | 6.8 ± 0.2 | 11.8 ± 0.3 | 15.4 ± 0.3 |

[a]Sensitivity to 200 °C blackbody radiation.
[b]Calculated using Eq. (9) with $R = R_{tot}$.

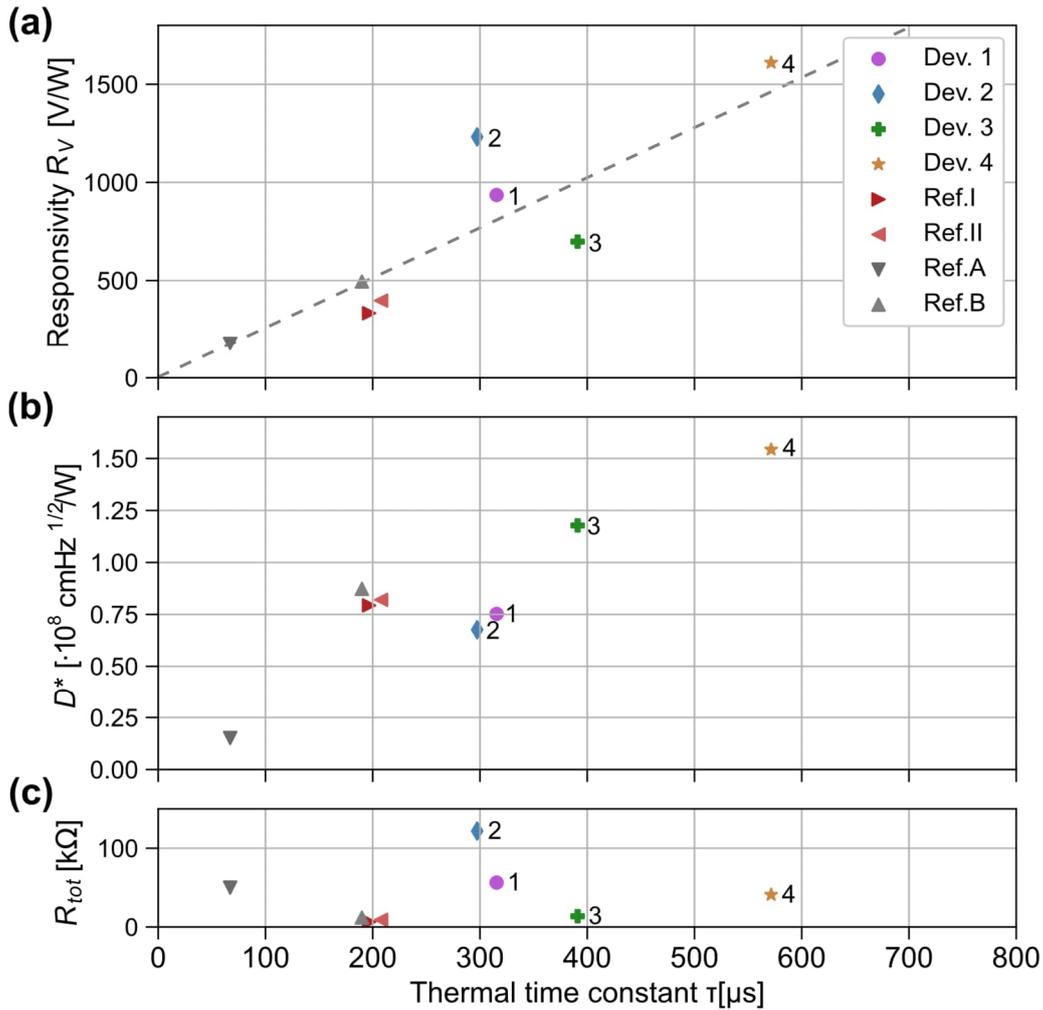

Figure 8. Graphs to compare (a) responsivities $R_V$, (b) specific detectivities $D^*$, and (c) total resistances $R_{tot}$ of the devices in this work and previously published devices. Reference devices I-II and A-B are from Refs. [21] and [14], respectively. Each device geometry in the figure is depicted with a different shape for easier comparison. A straight line running from origin to the Dev. B datapoint has been added to the responsivity graph to aid in the comparison of the devices.

The devices presented in this work (Devs. 1-4) show in general greater responsivities, but larger thermal time constants compared to the devices with a thicker poly-Si thermoelectric layer, owing to the thickness scaling as described by Eqs. (5) and (6). Furthermore, these poly-Si membranes are expected to have lower thermal conductivity which increases the responsivity further [see Eq. (5)]. This is indicated by the data of Table 2 and Fig. 8(a) as all the 40 nm devices exhibit the highest responsivities. The thickness reduction also leads to lower heat capacity which reduces the increase of the thermal time constant $\tau$. This effect of increased thermal isolation can be compensated in the device design by widening the thermoelectric beams. To employ the model of Eq. (4), a straight line running from origin to the Dev. B datapoint has been added into Fig. 8(a) to aid in the comparison of the devices. Due to the similarity of the doping levels all the devices in Fig. 8 have similar Seebeck coefficient $S$ values. The absorptance values estimated by effective absorber impedances and optical simulations are also similar for Devs. A, B, 1, and 4, allowing us to employ Eq. (4) in the device comparison. In Fig. 8(a), Devs. 1 and 4 are above this line, indicating that they have smaller heat capacities than Dev. A according to Eq. (4). This agrees with the thinner thermoelectric and absorber layers of Devs. 1-4. When comparing devices of the same geometry in this work (Dev. 2) to the previous works (Devs. A and B), there is a clear improvement in the responsivity-speed trade-off as Dev. 2 stands out well above the fitted visualization line.

Compared to previously published data from devices with thicker poly-Si [14,21], the decrease in poly-Si thickness from 80 nm to 40 nm resulted in an increase in film resistivity from 4.7 mΩcm and 3.0 mΩcm to 8.0 mΩcm and 4.9 mΩcm for n- and p-type poly-Si, respectively. The increased film resistivity in conjunction with thinner film thickness results in approximately a 3-fold increase in sheet resistance of the thinner poly-Si devices. The charge-carrier mobility decreases from 24 cm$^2$/Vs and 14 cm$^2$/Vs to 8.9 cm$^2$/Vs and 7.6 cm$^2$/Vs for n- and p-type poly-Si, respectively, due to increased surface scattering from the reduced film thickness, and in the case of the n-type poly-Si increased dopant concentrations from 5.5 x 10$^{19}$ cm$^{-3}$ to 8.8 x 10$^{19}$ cm$^{-3}$ owing to increased implantation dose.

To estimate the thermal material properties of the thinner poly-Si membranes we use Eq. (3) with literature estimates of the Seebeck coefficients of the n- and p-type material derived from the data on LPCVD polycrystalline silicon with similar resistivities [49,50]. Literature data suggest that the n- and p-type poly-Si Seebeck coefficients are -0.25 mV/K ... -0.30 mV/K and 0.20 mV/K ... 0.25 mV/K, respectively. The total Seebeck coefficient, $S$, of the detectors is thus 0.45 mV/K ... 0.55 mV/K, which is in line with previously published data [14,21]. We will estimate the thermal conductivity for Dev. 2, as this detector has the same grid and thermoelectric beam dimensions as the devices discussed in Ref. [14]. Using Eq. (3) for Dev. 2 with the 200°C value of $R_v$ = 1252 V/W and the simulated total optical efficiency in the experimental conditions, $\eta_{tot}$ = 0.82, gives $S/G_{th}$ = 1502 V/W. The thermal conductance, $G_{th}$, of Dev. 2 can be estimated as 0.30...0.37 µW/K. By assuming the thermal conductivities of n- and p-type are equal we can estimate the thermal conductivity of the 40 nm poly-Si as 9...11 W/(mK) suggesting that this 40 nm poly-Si has smaller thermal conductivity than 80 nm poly-Si, which has estimated thermal conductivity the range of 17...21 W/(mK) [14]. These estimates are in line with the thermal conductivity values of single- and poly-crystalline silicon membranes with similar thicknesses [19,23,39,51,52]. These estimates are considerably lower than the thermal conductivity of 40 nm thick single-crystalline Si nanomembrane, which can be estimated to be 25 W/(mK) using a phenomenological model [41] based on experimental data [23]. Reducing the thermal conductivity of silicon is one way to increase $ZT$ and thus improve its performance as a thermoelectric material. The low thermal conductivity of the poly-Si membranes suggests that the spatial confinement of phonon modes and increased boundary scattering reduce the thermal conductivity of the membrane [28,31,32].

The wavelength-dependent response characteristics of the devices simulated using the characteristics of Dev. 2 as reference are shown in Fig. 9(a). The spectral optical efficiency and responsivity are relatively flat over the 8–10 µm range, with a peak close to 10 µm corresponding with the peak absorptance of the quarter-wave resistive absorber (see Section 2.2). The periodic narrower absorption peaks at shorter wavelengths exhibit further characteristics of a quarter-wave resistive absorber. The calculated peak responsivity is slightly greater than the measured experimental value due to the experimental measurement considering the total optical efficiency, $\eta_{tot}$, over the whole spectral range of the incident radiation, rather than just the wavelength-dependent optical spectral efficiency, $\eta(\lambda)$. The experimental spectral radiance on the detector and the spectral absorptance are shown in Fig. 9(b). They indicate that during characterization the majority of the optical power absorbed in the detector is the wavelength range 5–11 µm.

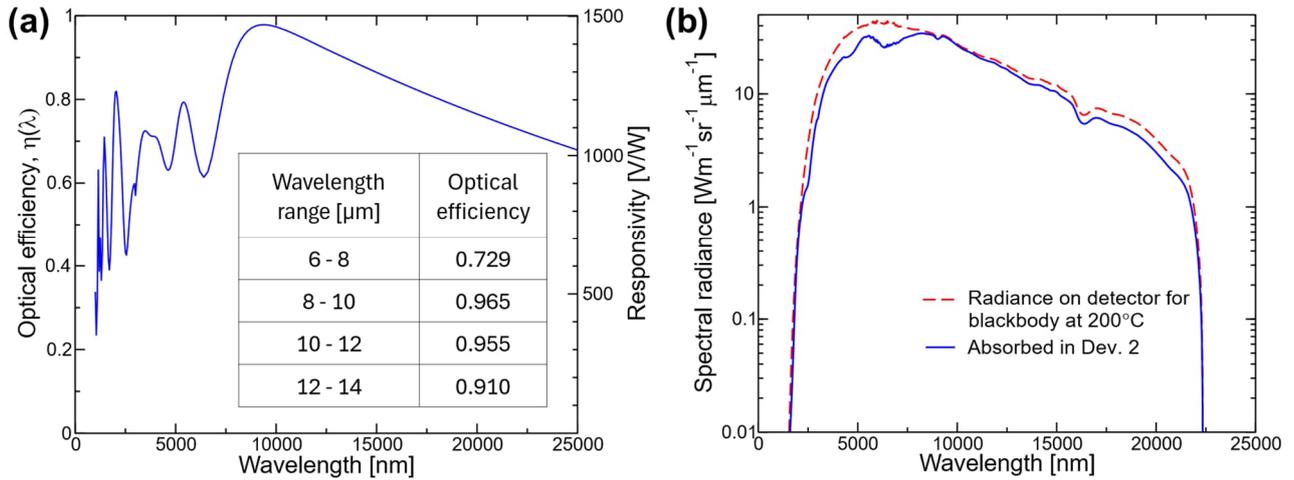

Figure 9. The response characteristics of the devices calculated using Dev. 2 as reference. Simulated spectral (a) optical efficiency and responsivity $R_V$, and (b) incident and absorbed radiance on the device during the experimental characterization. Inset in (a) are tabulated values of the optical efficiencies for specific wavelength ranges. The responsivity $R_V$ in (a) was calculated using $R_V(\lambda) = S\eta(\lambda)/G_{th}$ with $S/G_{th}$ = 1502 V/W.

To study the factors affecting the response time characteristics of the devices, we plot Eq. (6) against key device design and material parameters, shown in Fig. 10. From Eq. (6) it can be seen that the number squares in the n- and p-type thermoelectric elements, $N$, and the total area occupied by the absorber material, $A_{abs}^{mat}$, are linearly proportional to device speed, $\tau$. The dependency of device speed on the thermoelectric layer thickness, $t_{TE}$, and specific heat capacity of the thermoelectric layer, $c_{V,TE}$, in the case of Device 2 is shown in Fig. 10(a). Decreasing $c_{V,TE}$ increases device speed due to reduced thermalization times. Reducing $t_{TE}$ reduces the heat capacity and the thermal conductance of the device, resulting in a net effect of slower device speed. For the same reason, reduction in $t_{TE}$ also increases responsivity [see Eq. (5)]. The increase in responsivity due to decreased $t_{TE}$ is much larger than the loss of device speed – decreasing $t_{TE}$ from 80 nm to 40 nm results in a ~36% increase in time constant for a ~100% increase in responsivity. The reference specific heat capacity for silicon used is the bulk silicon value $c_{V,TE}$ = 1.64·10⁶ J/K/m³. It should be noted that the specific heat capacity for membranes can be up to several times greater than the bulk value [53], which explains the discrepancy between the calculated and experimental thermal time constants for Dev. 2. Scaling down $t_{abs}$ reduces the time constant of the device, shown in Fig. 10(b). The absorber material thickness $t_{abs}$ has no direct effect on responsivity if optimal impedance matching can be achieved by optimizing absorber grid geometry. Optimal device performance is thus achieved with minimal $t_{abs}$, within fabrication and process integration limits.

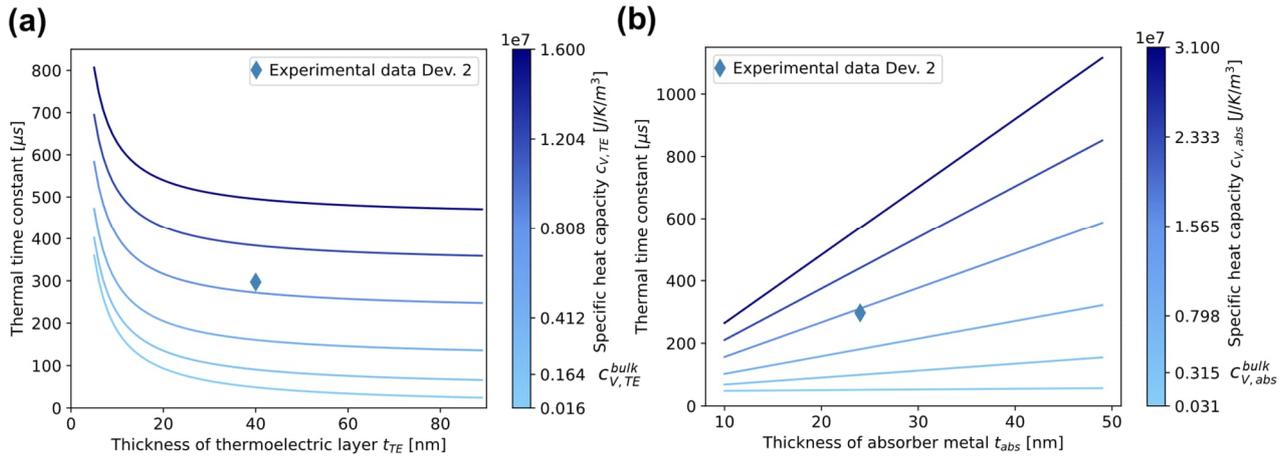

Figure 10. Analytical dependence of device thermal time constant on the thickness and specific heat capacity of the (a) thermoelectric layer and (b) absorber metal, calculated using Eq. (6) and the characteristics of Dev. 2 as reference. The specific heat capacity is varied within two orders of magnitude of the bulk value. The experimental datapoint for Dev. 2 is included for reference. The thickness of the absorber metal $t_{abs}$ is 24 nm in (a) and $t_{TE}$ is 40 nm in (b). The value of thermal conductivity is $\kappa_p = \kappa_n = 8.84$ W/(mK).

The reduced thermal conductivity from reducing the thermoelectric layer thickness results in increased $ZT$, as the increase in electrical resistivity is comparatively smaller. However, despite having high responsivity, the devices show relatively low detectivities, shown in Fig. 8(b). This is due to the relatively large and varying electrical resistance of the devices, shown in Fig. 8(c). The total electrical resistance $R$ determines the intrinsic voltage noise of the detector, and therefore directly affects detectivity. The electrical resistance of the detector comprises both the series resistance of the thermoelectric transducers and the poly-Si-TiN contact resistance. Here we define the total measured resistance of the bolometer as $R_{tot}$. By decreasing the thickness of the poly-Si layer, both the conductive thickness and electrical resistivity are affected. This results in an approximately a 3-fold increase in sheet resistance of the thinner poly-Si devices. Despite this, for most of the devices the dominating contribution to $R$, and therefore the main reason for decreased detectivity, is contact resistance. For Devs. 1-2, the resistance arising from the thermoelectric beams is $R_{beams} = 7.6$ $k\Omega$ and for Devs. 3-4 this contribution is $R_{beams} = 11.5$ $k\Omega$. Particularly for Devs. 1 and 2 the total resistances of the thermoelectric transducers are $R_{tot} = 56.3$ $k\Omega$ and $R_{tot} = 121.8$ $k\Omega$, respectively. Ideally, the total bolometer resistance would be determined by the beams only, i.e. $R_{beams}$. In this case the NEPs and specific detectivities of Devs. 1-4 would reach 8...19 pW/ Hz$^{1/2}$ and 1...3·10$^8$ cmHz$^{1/2}$/W, respectively.

The high and varying contact resistance is caused by two main factors: a lower quality poly-Si film (in terms of topography) and HF-vapor (HFV) etching damage to the poly-Si-TiN contacts due to the rough contact interface. The effect of HFV on damaging the contact resistance is additionally affected by geometry. Using the contact resistivity measured by CBK (4×4μm$^2$), the contact resistances of Devs. 1 and 2 can be estimated to be only $R_{contact} = 82$ $\Omega$ and $R_{contact} = 96$ $\Omega$, respectively. In the CBK test structure the relatively large contact area of 4×4 μm$^2$ is largely covered and protected against HFV, whereas the grids with the maximum linewidths of ~0.5–1.2 μm are more exposed. The effect of high $R$ can be seen to decrease the detectivities of Devs. 1 and 2 in Fig. 8(c). This effect can also be seen to a lesser extent to decrease the detectivities of Devs. 4 and A [14] and to result in a moderate detectivity for Dev. 3.

In reducing the deposited poly-Si thickness to 40 nm, the film uniformity suffers as the LPCVD deposition process was not optimized for ultra-thin poly-Si film deposition. The deposited film exhibits a large number of island-type growth bumps that act to increase the total film roughness and contribute periodic topographic irregularities in the poly-Si-TiN contact area (see Fig. S1 in Supplementary Information for full AFM scans). The defect density of ~40 nm growth-bumps in the 40 nm poly-Si film increased to 8.6 defects/$\mu m^2$ from 1.9 defects/$\mu m^2$ from devices reported in previous works [14,21]. When combined with the reduced TiN thickness of 24 nm compared to 50 nm in previously published devices, the poly-Si-TiN interface is more susceptible to contact damage during the HF vapor release step. The contact resistance in general is highly sensitive to the poly-Si-TiN interfacial roughness during the HF vapor etching process, which can cause variations in $R_{\text{tot}}$. In future work we plan to resolve this growth limitation with improved ultra-thin poly-Si deposition techniques capable of topography-free films down to 10 nm. Furthermore, the HFV process can be optimized to reduce contact resistance degradation.

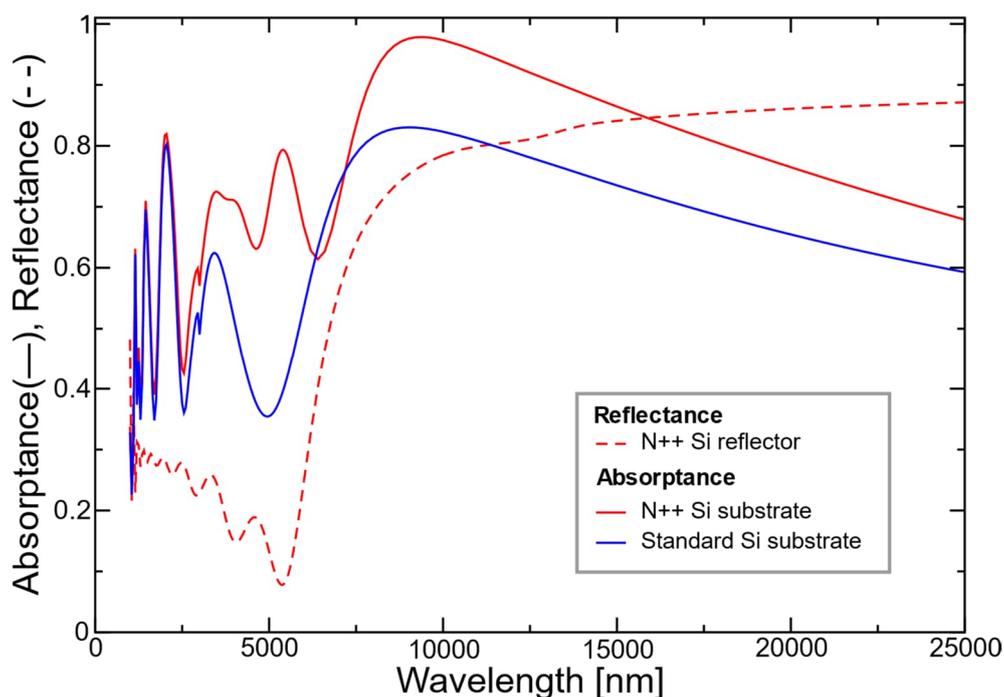

Figure 11. Simulated absorptance of Dev. 2 with an N++ surface doped back reflector and standard Si substrate. The dopant concentration of the N++ Si substrate is $10^{20}$ cm$^{-3}$. The back reflector increases peak absorptivity by ~20%. The reflectance of the N++ mirror is included for reference.

Another way to improve the performance of the devices is to improve the reflectance of the back reflector, which is in this case the Si substrate. Devs. 2-3 had additional surface processing to form an N++ surface doped back reflector on the silicon substrates. Fig. 11 shows the simulated spectral absorptance of Dev. 2 with the N++ surface doped back reflector compared to a standard Si substrate. It should be noted that the simulation yields slightly higher peak absorptance than the semi-analytical model for Dev. 2. The N++ surface increases the reflectance of the Si substrate from approximately 30% to 70%–75% for wavelengths between 8–10 µm. This is in line with our experimentally measured reflectance of the N++ surface doped substrate of 70% and 75% at 8 µm and 10 µm, respectively. In the case of Devs. 2-3, this increase in reflectance translates to a ~20% increase in the estimated peak absorptivity. The effect of the N++ doped back reflector can be seen when comparing the responsivities of Devs. 1 and 2, which both have close to ideal impedance matching and identical thermoelectric beam dimensions (see Table 1). The responsivity of Dev. 2 (1252 V/W) is about 30% larger than Dev. 1 (951 V/W), due to the inclusion of the N++ doped back reflector, minor differences in device impedance and grid geometry. In the case of Dev. 3 this effect is not strongly visible due to poor impedance matching.

# 5   Conclusions

In conclusion, we demonstrated LWIR nano-thermoelectric bolometers based on ultra-thin (40 nm) poly-Si nanomembranes and integrated back-reflectors. These bolometers utilize CMOS-compatible materials and fabrication, and their performance is suitable to a wide range of applications by tailoring the speed and sensitivity of the devices by controlling device dimensions. We report devices with responsivities up to 1636 V/W with time constants down to 297 µs. The integrated back-reflector was formed by a heavily doped (N++) layer on Si substrate, which led to a ~20% increase in the estimated peak absorptivity, translating to ~30% increase in responsivity, in comparison to an undoped substrate. The high sensitivity and high-speed operation stems from the low thermal conductance and low heat capacity device structure. Here, the thermoelectric beams act both as mechanical supports and transducer elements providing low thermal conductance to the absorber and eliminating the need for any additional micromechanical supports. A single metal layer grid functions as both the absorbing element for incident radiation and an electrical connection between the poly-Si n- and p-type thermoelectric elements, thereby introducing minimum additional heat capacity. Due to contact resistance issues in the devices arising from rough poly-Si-TiN interface and fine device geometry, the specific detectivities of the devices reported here are limited. By resolving the contact resistance issue so that the dominating component of device resistance is the thermoelectric beam resistance we estimate specific detectivities up to $3 \cdot 10^8$ cmHz$^{1/2}$/W and NEPs down to 19 pW for the devices discussed in this work. In future work we plan to resolve the contact resistance issue with improved ultra-thin poly-Si deposition techniques capable of topography-free films down to 10 nm, and by optimizing the HF vapour release process to reduce poly-Si-TiN contact resistance degradation.

## Supplementary Material

See the supplementary material for AFM data on poly-Si topography and supporting details on IR characterization and optical modelling of absorbers.

## Acknowledgements

We acknowledge gratefully the technical assistance of Teija Häkkinen in device fabrication. This work has been financially supported by European Union Future and Emerging Technologies (FET) Open under Horizon 2020 programme (Grant Agreement No. 766853, project EFINED), by Business Finland co-innovation projects RaPtor and HigPlg (Nos. 6030/31/2018 and 4380/31/2023), and by The Research Council of Finland (Grant No. 342586). The work of Jonna Tiira was supported by The Research Council of Finland through Grant No. 324838. This work is part of the Research Council of Finland Flagship Programme, Photonics Research and Innovation (PREIN), decision 320168.

## Data availability

The data that supports the findings of this study are available within the article and its supplementary material.

# Supplementary information: Infrared Bolometers Based on 40-nm-Thick Nano-Thermoelectric Silicon Membranes

Anton Murros [a)], Kuura Sovanto, Jonna Tiira, Kirsi Tappura, Mika Prunnila, and Aapo Varpula[a)]

VTT Technical Research Centre of Finland Ltd, Tietotie 3, FI-02150 Espoo, Finland
[a)] Authors to whom correspondence should be addressed: anton.murros@vtt.fi and aapo.varpula@vtt.fi

## 1 AFM scans of topological defects in 40 nm poly-Si

Atomic force microscopy scans of 40 nm n- and p-type poly-Si membranes before sacrificial oxide etching for membrane suspension are shown in Fig. S1. The high defect density resulting from unoptimized LPCVD ultra-thin film growth can be seen as an increased number of topographic defects when compared to 80 nm poly-Si described in previous work [1]. The defect density of ~40 nm growth-bumps in the 40 nm poly-Si film increased to 8.6 defects/$\mu m^2$ from 1.9 defects/$\mu m^2$.

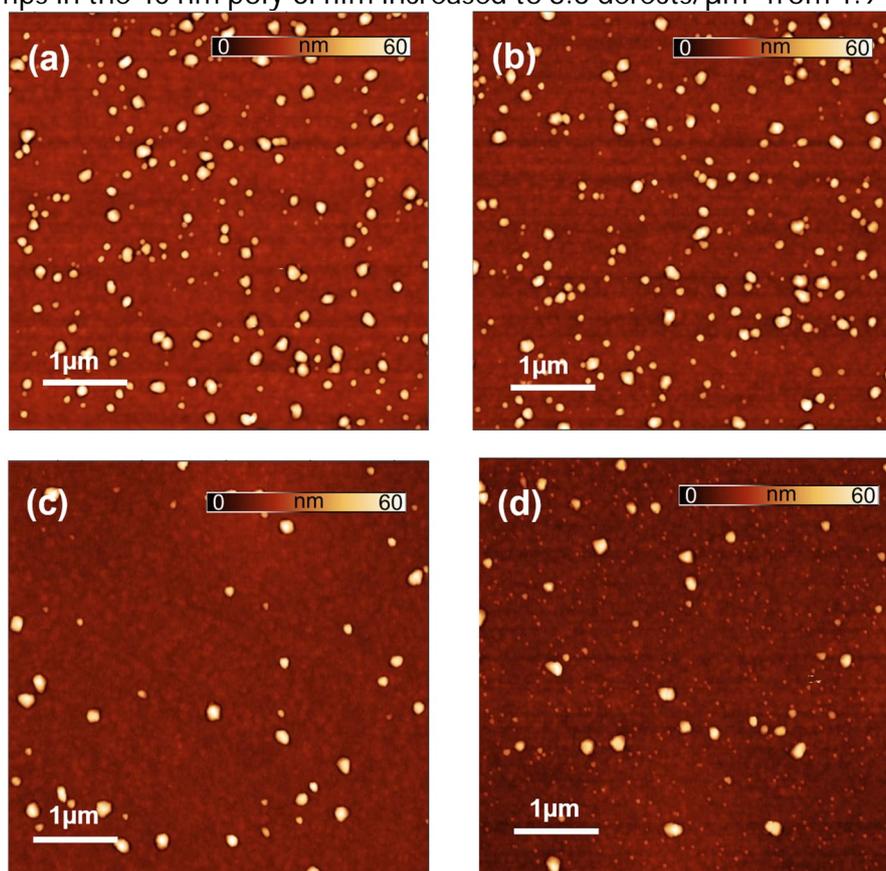

Supplementary figure S1. Atomic force microscopy (AFM) scans of supported 40 nm poly-Si a) n-type b) p-type membranes before release. Reference AFM scans from our previous work (see Ref. [1]) with 80 nm poly-Si c) n-type and d) p-type membranes before release. Scale bar for all scans is 60nm.

## Supplementary references